\documentclass[11pt,a4paper]{article}

\usepackage[utf8]{inputenc}
\usepackage{amsmath}
\usepackage[mathcal]{euscript}
\usepackage{latexsym}
\usepackage{textcomp}
\usepackage{slashed}
\usepackage{tikz}
\usepackage{xspace}
\usepackage{setspace}
\usepackage[a4paper]{geometry}
\usepackage[numbers,sort&compress]{natbib}
\newcommand{\SAMURAI}{{\textsc{Sa\-mu\-rai}}}

\newcommand{\bea}{\begin{eqnarray}}
\newcommand{\eea}{\end{eqnarray}\noindent}

\newcommand{\beq}{\begin{equation}}
\newcommand{\eeq}{\end{equation}}
\def\ff{f\hspace{-0.3cm}f}

\newcommand{\bcen}{\begin{center}}
\newcommand{\ecen}{\end{center}}
\def\url#1{\texttt{#1}}

\def\Gosam{{{\sc GoSam}}}
\def\gosam{{{\sc GoSam}}}
\def\gosamtwo{{{\sc GoSam 2.0}}}

\def\C++{{{\sc c++}}}

\def\Golem{{{\sc Golem95C}}}

\def\Ninja{{{\sc Ninja}}}

\def\ttw{{$pp \to t \bar t W$}}
\def\ttz{{$pp \to t \bar t Z$}}
\def\tth{{$pp \to t \bar t H$}}

\begin{document}

\begin{center}
%\begin{spacing}{1.5}

{\Large\bf Recent Developments in Higher-Order Calculations: \\ \medskip Hard Functions at NLO with \Gosam} \\
%\end{spacing}
\vskip 9mm

{\bf Alessandro Broggio}$^a$\\
\small{\tt alessandro.broggio@tum.de}
\\[0.5 em]

{\bf Andrea Ferroglia}$^{b,c}$\\
\small{\tt aferroglia@citytech.cuny.edu}
\\[0.5 em]

{\bf Nicolas Greiner}$^{d}$ \\
\small{\tt greiner@physik.uzh.ch}
\\[0.5 em]

{\bf Giovanni Ossola}$^{b,c,}$\footnote{Speaker}\\
\small{\tt gossola@citytech.cuny.edu}
\\[0.5 em]

\vskip 2mm

{\small 
{\it $^a$Physik Department T31, Technische Universit\"at M\"unchen, \\
James Franck-Stra{\ss}e 1, D-85748 Garching, Germany} \\
{\it  $^b$New York City College of Technology, 
  City University of New York, \\ 300 Jay Street, Brooklyn NY 11201, USA} \\
{\it $^c$The Graduate School and University Center, 
  City University of New York,  \\ 
  365 Fifth Avenue, New York, NY 10016, USA}\\
{\it  $^d$Physik Institut, Universit{\"a}t Z{\"u}rich, \\ Winterthurerstr.190, 8057 Z\"urich, Switzerland  }\\
}
\end{center}

\vspace{0.2cm}

\begin{abstract}
In this presentation, we will focus on a recent applications of the \gosamtwo\ automated framework which goes beyond the official scope of this code, originally designed for one-loop fixed order calculations. In particular, we describe a customized version of \Gosam\ that has been recently employed to study the production of a top-antitop pair in association with a vector boson or with the Higgs boson to next-to-next-to-leading logarithmic accuracy. In the context of these calculations, the modified version of \gosam\ was used to evaluate the NLO hard functions which are needed to carry out the resummation of soft gluon emission effects. We also briefly comment on the ongoing efforts to generalize integrand reduction and unitarity for higher order calculation, towards the goal of developing efficient alternative computational techniques for the evaluation of scattering amplitudes beyond one loop. 
\end{abstract}

\vspace{0.2cm}

\begin{center}
%\begin{spacing}{1.5}
{\it to appear in the proceedings of } \\

\vspace{0.2cm}

The European Physical Society Conference on High Energy Physics\\
		5-12 July 2017,  Venice, Italy
%\end{spacing}
\end{center}

\newpage

\section{Introduction}

Automation for one-loop calculation has been successfully achieved over the past decade. Thanks to mathematical and technical development in the field, several algorithms have been designed and implemented in a series of highly efficient multi-purpose computational tools. %As a consequence, just like the case of tree-level processes, most next-to-leading order (NLO) calculations can be efficiently done by preparing an input card and pushing a button (and eventually waiting to achieve enough statistics). 

The situation beyond one loop is much more variegated. The theoretical understanding of the structure of scattering amplitudes at multi-loop has been the subject of several studies. 
%Using the language , such as the case of algebraic geometry, led to cohesive picture in the description of multi-loop amplitudes and to several ways of extending and generalizing the integrand reduction approach to multi-loop amplitude in a highly non-trivial manner. Rigorous proofs, such as the \emph{Maximum-Cut Theorem} showed that the construction of amplitudes on their kinematic cuts can be safely achieved for any number of loops. 
The use of new mathematical approaches, in particular techniques borrowed from algebraic geometry, led to elegant and general results for the structure of the integrands of scattering amplitudes. Moreover, by means of rigorous proofs, such as the \emph{Maximum-Cut Theorem}, it was shown that the construction of amplitudes based on their kinematic cuts can be safely achieved for any number of loops.  Overall, on the one hand, the progress in the field is undeniable. On the other hand, the conceptual progress has not yet been translated in efficient new computational algorithms. At present, calculations beyond one loop are for the most part performed following the ``traditional'' path of generating Feynman diagrams with computer algebra, reducing them by tensorial reduction and projections over convenient form factors, minimizing the number of Master Integrals (MIs) using integration-by-parts (IBP) identities, and finally evaluating a minimal set of MIs using analytic expressions if possible, otherwise numerically. The hope for the near future is that, as already realized at one-loop, also two- and higher-loop calculations will benefit from the development of new algorithms based on unitarity or higher-order integrand reduction.

In this conference paper, we will report on two different projects, that share the common feature of going beyond the one loop level. In the first and more extensive part, we will discuss an extension of the \Gosam\ framework for automated one-loop calculations~\cite{Cullen:2011ac, Cullen:2014yla} to evaluate the NLO hard functions which are needed to carry out the resummation of soft gluon emission effects. In a series of recent papers~\cite{Broggio:2015lya, Broggio:2016zgg, Broggio:2016lfj, Broggio:2017kzi}, this novel feature of \Gosam\ was applied to study the associated production of a top quark pair and a boson (Higgs, $W$, or $Z$) at next-to-next-to-leading logarithmic (NNLL) accuracy. A second topic, namely an overview of the recent developments towards an integrand-reduction based approach to two-loop (and higher-order) calculations, will be the subject of our final outlook.

\section{Soft Limit and Factorization} \label{soft}

The partonic cross section for top pair production in association with a Higgs, $W$, or $Z$ boson
receives potentially large corrections from soft gluon emission diagrams.
Schematically, the partonic cross section depends on logarithms of the ratio of two different scales, 
$$L=\ln{\left( \frac{\text{``hard'' scale}}{\text{``soft'' scale}} \right)}$$
where potentially $L \alpha_s \sim 1$. In this situation, the powers of $\alpha_s$ alone do not guarantee the correct hierarchy between the various terms in the expansion, and one needs to reorganize the perturbative expansion and resum these large corrections to all orders in perturbation theory.

The resummation of these effects to NNLL accuracy can be carried out by exploiting the factorization properties of the partonic cross section in the soft limit, which can be studied with effective field theory methods~\cite{Becher:2014oda} and by subsequently  employing renormalization group improved perturbation theory techniques. In the following, for simplicity, we summarize the main formulas in the case of \ttw\ production. Analogous formulae can be 
obtained for \ttz\ and \tth, we refer the reader to~\cite{Broggio:2015lya, Broggio:2016zgg, Broggio:2016lfj, Broggio:2017kzi} for more details.

The associated production of a top quark pair and $W$ boson receives contributions from the partonic processes 
\begin{align}
q(p_1) +  \bar q(p_2) \longrightarrow t(p_3) +\bar{t} (p_4) + W(p_5)  + X \, , 
\label{eq:partproc}
\end{align}
where  $X$ indicates the unobserved partonic final-state radiation.
After defining the invariants $\hat{s} = (p_1+p_2)^2  = 2 p_1 \cdot p_2 $ and $M^2 = \left(p_3 + p_4 + p_5 \right)^2$, we define the soft or partonic threshold limit as the region in which  $z \equiv M^2/\hat{s} \rightarrow 1$. Indeed, in this kinematic region, the final state radiation can only be soft.
We write the factorization formula for the cross section in the partonic threshold limit as 
{\small \beq
\sigma \left(s,m_t,m_W \right) = \frac{1}{2 s} \int_{\tau_{\text{min}}}^{1} \!\!\! d \tau \int^1_{\tau} \frac{dz}{\sqrt{z}} \sum_{ij} \ff_{ij} \left( \frac{\tau}{z}, \mu \right)  \int d\text{PS}_{t\bar{t}W} \mbox{Tr}\left[\mathbf{H}_{ij}\left(\{p\},\mu\right) \mathbf{S}_{ij}\left(\frac{M (1-z)}{\sqrt{z}},\{p\},\mu\right)  \right] \, ,
\label{eq:factorization}
\eeq}
where $s$ the square of the hadronic center-of-mass energy,  $\tau_{\text{min}} = \left(2 m_t + m_W\right)^2/s$, and $\tau = M^2/s$.
Following~\cite{Broggio:2015lya, Broggio:2016zgg, Broggio:2016lfj, Broggio:2017kzi}, we denote the hard functions with $\mathbf{H}$,
the soft functions with $ \mathbf{S}$, and the luminosity functions with
$\ff$. We refer the reader to these papers for more details.  

For $q\bar{q}$-initiated processes, such as \ttw, the hard and soft functions are two-by-two matrices in color space (these become three-by-three matrices in color space for $gg$-initiated processes, which appear for example in \ttz). The hard functions satisfy renormalization group equations governed by the soft anomalous dimension matrices $\Gamma^{ij}_H$~\cite{Ferroglia:2009ep,Ferroglia:2009ii}. In order to carry out the resummation to NNLL accuracy, the hard functions, soft functions, and soft anomalous dimensions must be computed at NLO accuracy.
In~\cite{Broggio:2016zgg, Broggio:2016lfj, Broggio:2017kzi}, this resummation program was carried out in Mellin space.

The NLO soft functions and soft anomalous dimensions are the same for the three processes discussed above, and are provided in~\cite{Broggio:2015lya,Broggio:2016lfj,Broggio:2016zgg}. On the contrary, NLO hard functions are process dependent and receive contributions exclusively from NLO virtual corrections. Such objects can be indeed computed by means of customized version of the automated tools for NLO calculations. %IN the following, we describe briefly how this goal has been achieved within {\tt GoSam}~\cite{Cullen:2011ac, Cullen:2014yla,Binoth:2008uq,Mastrolia:2010nb}. 

\section{Hard functions at NLO with GoSam} \label{hard}

The {\gosam} framework~\cite{Cullen:2011ac, Cullen:2014yla} combines automated Feynman diagram generation~\cite{Nogueira:1991ex, Vermaseren:2000nd, Reiter:2009ts, Cullen:2010jv}, with a variety of reduction techniques, to allow for the automated numerical evaluation of virtual one-loop 
%QCD or electroweak correction~\cite{} 
corrections to any given process. After all relevant Feynman integrals are generated, virtual corrections can be evaluated using the integrand reduction via Laurent expansion~\cite{Mastrolia:2012bu,vanDeurzen:2013saa} provided by {\Ninja}~\cite{Peraro:2014cba}, the $d$-dimensional integrand-level reduction method~\cite{Ossola:2006us,Mastrolia:2008jb,Ossola:2008xq}, as implemented in \SAMURAI~\cite{Mastrolia:2010nb}, or the tensorial decomposition provided by {\Golem}~\cite{Binoth:2008uq,Heinrich:2010ax, Guillet:2013msa}. 

By default, \gosam\ computes squared amplitudes summed over
colors. To build the hard functions we need instead to combine color
decomposed amplitudes. 

%In the models used by \gosam{}, we allow one unbroken gauge group
%$\mathsf{SU}(N_C)$ to be treated implicitly; any additional gauge group,
%broken or unbroken, needs to be expanded explicitly. Any particle
%of the model may be charged under the $\mathsf{SU}(N_C)$ group in the
%trivial, (anti-)fundamental or adjoint representation. Other representations
%are currently not implemented.
For each given process, \gosam\  projects each Feynman diagram onto an appropriate process dependend color basis $\left\{ | c_{i}\rangle \right\}$, thus casting it in the form
\begin{equation}
\mathcal{D}=\sum_{i=1}^k \mathcal{C}_i \vert c_{i}\rangle\, .
\end{equation}
After the basis is determined, the code computes, once for all, all entries of the matrices
$\langle c_i\vert c_j\rangle$ %and $\langle c_i\vert T_I\cdot T_J\vert c_j\rangle$. 
which will be relevant for the evaluation of the hard functions. 
% The latter is used to provide color correlated Born matrix elements used to for check the
%IR poles of the virtual amplitudes.
%
%For the above example, \gosam{} obtains\footnote{%
%In the actual code the results are given in terms of $T_R$
%and~$N_C$ only.}
%\begin{equation}
%\langle c_i\vert c_j\rangle=T_R C_F\left(\begin{array}{ccc}%
%(N_C^2-1) & -1 & N_C \\
%-1 & (N_C^2-1) & N_C \\
%N_C & N_C & N_C^2
%\end{array}\right).
%\end{equation}
%Similarly, the program computes the matrices $\langle c_i\vert T_I\cdot T_J%
%\vert c_j\rangle$ for all pairs of partons $I$ and $J$.
For any given process, we denote the tree-level matrix element and the  tree-level squared amplitude respectively as 
\begin{equation}
| \mathcal{M}^{(0)} \rangle=\sum_{j=1}^k \mathcal{C}^{(0)}_j \vert c_j\rangle \quad  \quad  
\text{and}  \quad   \quad
\left\vert\mathcal{M}^{(0)}\right\vert^2=\sum_{i,j=1}^k
\left(\mathcal{C}^{(0)}_i\right)^\ast\mathcal{C}^{(0)}_j
\langle c_i\vert c_j\rangle \, . \label{tree}
\end{equation}
In order to obtain the NLO prediction, we should compute the interference term between the tree-level and one-loop matrix-elements.
We perform this contraction already at the integrand level
\begin{equation}
\mathcal{N}_\alpha(q)=\sum_{i,j=1}^k
\langle c_i\vert c_j\rangle
\left(\mathcal{C}^{(0)}_i\right)^\ast\mathcal{C}^{(1)}_j(q), \label{nlo}
\end{equation}
where $\mathcal{C}_j^{(1)}$ is formed by the sum over the corresponding
coefficients of all diagrams that share a set of denominators. The numerator functions obtained in this manner are then passed to the reduction.

In order to compute the hard functions, instead of summing over all colors, we need to extract all contributions to the virtual amplitudes for fixed values of $i$ and $j$. The LO and NLO hard function can be indeed evaluated as 
\begin{equation}  \mathbf{H}^{(0)}_{ij}   = \frac{1}{4} \frac{1}{\langle c_i | c_i \rangle \langle c_j | c_j \rangle} \langle c_i | 
\mathcal{M}^{(0)} \rangle \langle \mathcal{M}^{(0)} |c_j \rangle \end{equation}
and
\begin{equation}  \mathbf{H}^{(1)}_{ij} =  \frac{1}{4} \frac{1}{\langle c_i | c_i \rangle \langle c_j | c_j \rangle} \left[ \langle c_i | 
\mathcal{M}^{(1)} \rangle \langle \mathcal{M}^{(0)} |c_j \rangle + \langle c_i | 
\mathcal{M}^{(0)} \rangle \langle \mathcal{M}^{(1)} |c_j \rangle \right] \, ,
\end{equation}
respectively. 
It is important to observe that both formulas (\ref{tree}) and (\ref{nlo}) are written in terms of the products $\langle c_i\vert c_j\rangle$ of elements of the color matrix. Therefore, if we want to isolate one specific color contribution to the LO and NLO amplitudes as required by the evaluation of the hard functions, we can simply set to zero all elements  $\langle c_i\vert c_j\rangle$ but one and then loop over all possible elements. %In we can easily build the color decomposed amplitudes.

For this purpose,  \gosam\ has been endowed with the new extension ``{\tt hardfunction}''. By specifying this keyword in the  \gosam\ input card, together with the initial and final state particles to be considered, the user can force the code to compute all different elements of the hard function matrix by projecting on different elements of the color basis. All results are provided as complex numbers. Afterwards, a change of basis can be performed according to the user's choice. As an output, \gosam\ generates a FORTRAN routine that allows, for each given phase
space point, to compute the corresponding hard functions at LO and NLO. 

\section{Phenomenological motivation and applications}

The machinery discussed in Sections~\ref{soft} and~\ref{hard} was recently applied to study the associated production of a top quark pair and a boson (Higgs, $W$, or $Z$) at next-to-next-to-leading logarithmic (NNLL) accuracy~\cite{Broggio:2015lya, Broggio:2016zgg, Broggio:2016lfj, Broggio:2017kzi}.
Precise theoretical predictions for these processes have indeed a wide variety of phenomenological applications. 

The associated production of a top pair and a $Z$ or $W$
boson (\ttw\ and \ttz) are the two processes with the heaviest final states so far observed observed at the LHC. 
The study of \ttz\ provides direct access to the coupling of the top quark to the $Z$ boson. This would allow to distinguish the Standard Model (SM) prediction respect to several New Physics scenarios, that predict changes to this coupling with respect to the SM. The process \ttw, after considering the decay of the top quark, can lead to events with two leptons of the same sign in the final state, in combination with jets and missing energy.  These events, relatively rare in the SM, are useful within supersymmetry searches. Moreover, both \ttw\ and \ttz\ are relevant in the context of dark matter searches.

The search for events in which a Higgs boson is produced in association with a top-antitop  quark  pair  (\tth\
production)  is one experimental goals of Run II of the LHC. While the Standard Model cross section for this
process  is  quite  small,  its measurement would indeed  provide
important and direct  information on the Yukawa  coupling  of  the  Higgs  boson  to  the  top  quark, which  is
crucial for verifying the origin of fermion masses, and understanding the hierarchy
problem related to the mass of the Higgs boson. Moreover, its  precise measurement would place strict constraints on New Physics searches.  

\begin{figure}[h!]  
		\begin{center}
%		\begin{tabular}{c}
			\includegraphics[width=9.0cm]{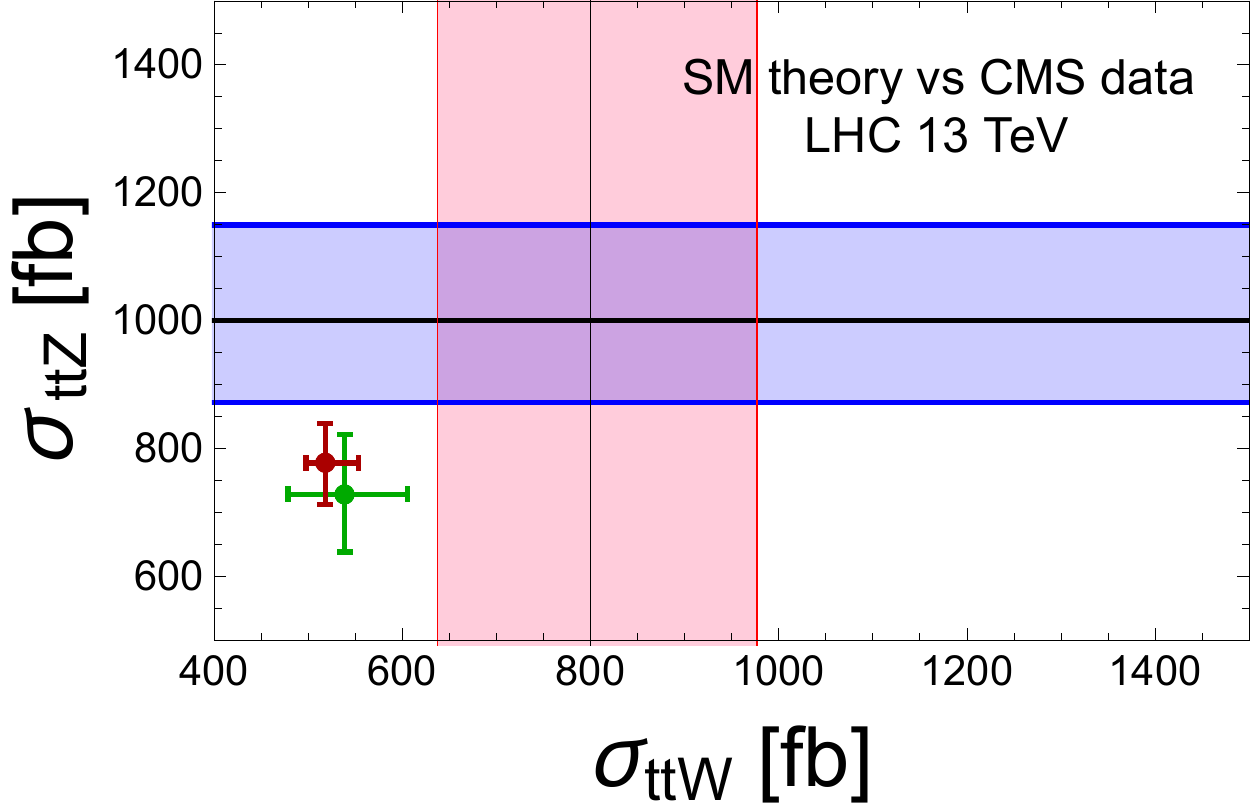} 
%		\end{tabular}
	\end{center} %\vspace{-0.6cm}
	\caption{Total cross section at NLO (green cross) and NLO+NNLL (red cross) compared to the CMS measurements at $13$~TeV~\cite{CMS:2017uib} (blue and pink bands).  \label{fig:totCS13}}
	\end{figure}

As an example of phenomenological application, in Figure~\ref{fig:totCS13} we present a comparison of the predictions for the total cross sections of \ttw\ and \ttz~\cite{ Broggio:2017kzi} at NLO and NLO+NNLL, with the corresponding experimental data for the LHC at $13$~TeV~\cite{CMS:2017uib} which were recently released by the CMS Collaboration. 

\vspace{0.3cm}

A previous version of this comparison, based on earlier data from ATLAS measurements at $8$~TeV~\cite{Aad:2015eua} and to the CMS measurements at $13$~TeV~\cite{CMS:2016dui}, showed that theoretical predictions for the $t \bar{t} W$ cross section were somehow smaller than measurements for both collider energies~\cite{ Broggio:2017kzi}. With the present data, the discrepancy has decreased well below the $2 \sigma$ limit, however the predicted SM values for both the $t \bar{t} W$ and $t \bar{t} Z$ cross sections remain
slightly smaller than the measured cross sections.  Of course, a fully exhaustive comparison between predictions and measurements should also account for the uncertainty associated to the choice of the PDFs and to the value of $\alpha_s$, which are not reflected in the error bars of Figure~\ref{fig:totCS13}.

The results were obtained by means of  an in-house parton level Monte Carlo code for the numerical evaluation of the resummation formula and matched to complete NLO calculations employing  \verb|MadGraph5_aMC@NLO|~\cite{Alwall:2014hca}.  The NLO hard function have been computed with the one-loop provider {\tt Openloops}~\cite{Cascioli:2011va} in combination with {\tt Collier}~\cite{Denner:2016kdg} and cross-checked with  {\tt GoSam}~\cite{Cullen:2011ac, Cullen:2014yla} used in combination with the reduction provided by {\tt Ninja}~\cite{Mastrolia:2012bu,vanDeurzen:2013saa,Peraro:2014cba}. 

\section{Future Outlook: towards a new approach to higher order calculations}

As anticipated in the introduction, we recently witnessed several interesting developments related to the mathematical properties of scattering amplitudes~\cite{Zhang:2016kfo}, which apply far beyond the well-known limit of one-loop calculations. Integrand reduction, revisited within the language of algebraic geometry, and unitarity started to build again that constructive interference pattern which led a few years back to revolutionary results for the one-loop case. 
While very few actual calculations appeared in the literature, an example being the two-loop four gluons amplitudes obtained with numerical unitarity method~\cite{Ita:2015tya, Abreu:2017xsl}, we expect these new approaches to play an increasingly important role in the forthcoming months.  

The idea of applying the integrand reduction beyond one loop, pioneered in~\cite{Mastrolia:2011pr, Badger:2012dp}, has been the target of several studies in the past five years, thus providing a new promising direction in the study of multi-loop amplitudes. An important upgrade in this process was achieved by systematizing the determination of the residues at the multiple poles of scattering amplitudes as a problem of multivariate polynomial division in algebraic geometry~\cite{Zhang:2012ce, Mastrolia:2012an,Mastrolia:2013kca}, which turned out to be a very natural language to describe the integrand-level decomposition. This approach confirms that the shape of the residues is uniquely determined by the on-shell conditions, without any additional constraint. 

The {\it Maximum Cut Theorem}~\cite{Mastrolia:2012an} guarantees that one of the pillars of integrand reduction, namely the construction and evaluation  of all residues on the kinematic cuts, is well grounded. After labeling as {\it Maximum-cuts} the largest sets of denominators which can be simultaneously set to zero for a given number of loop momenta, the {\it Maximum Cut Theorem} ensures that the corresponding residues are parametrized by exactly $n_s$ coefficients, where $n_s$ is the number of solutions of the multiple cut-conditions. This theorem extends to all orders the features of the one-loop quadruple-cut in dimension four~\cite{Britto:2004nc,Ossola:2006us}.

In a different and very promising approach to integrand reduction, called Adaptive Integrand Decomposition~\cite{Mastrolia:2016dhn}, non-physical degrees of freedom are integrated out by means of orthogonality relations, thus eliminating spurious integrals and leading to much simpler expressions for the integrand-decomposition formulae. 

While many computer algebra systems use finite fields for solving problems such as polynomial factorization, the application of finite fields in high-energy physics is very recent~\cite{vonManteuffel:2014ixa,Peraro:2016wsq, Maierhoefer:2017hyi}. Multivariate polynomials and rational functions, which commonly appear in the techniques mentioned above, can indeed be analytically reconstructed from their numerical evaluation at several values of its arguments, and finite field provide an extremely efficient way of achieving this goal.  

The \gosam\ framework was developed to compute one-loop virtual contributions, as needed by NLO phenomenology. However, thanks to its modular structure, it can be extended to address specific tasks needed by higher order calculations: for example, to produce expressions for the two-loop Feynman diagrams contributing to any given process. This feature has been successfully employed to compute the two-loop virtual amplitudes for Higgs boson pair production in gluon fusion~\cite{Borowka:2016ehy, Borowka:2016ypz, Heinrich:2017kxx}.% where all unknown MIs have been computed numerically by means of SecDec-3.0~\cite{Borowka:2015mxa}. 

To summarize, plenty of activities are currently in progress beyond one-loop. Traditional tools for IBP-based approaches have been upgraded with smarter techniques (finite fields reconstruction). Further improvements have been achieved in the analytic and numerical evaluation of Master Integrals.
Algebraic geometry provided a new handle to understand the structure of mathematical objects which appear in these calculations. Unitarity at higher loops is under development. Overall, it's a time of theoretical improvements and mathematical explorations, which will naturally lead to alternative approaches for advanced multi-loop calculations. 

{\small \paragraph{Acknowledgments}
%$\clubsuit$ to be UPDATED $\clubsuit$ 
%Results presented in this paper were obtained using the computer
%cluster of the Center for Theoretical Physics at the Physics Department of New
%York City College of Technology.
The research of A.B. is supported by the DFG cluster of excellence ``Origin and Structure of the Universe''. N.G. was supported by the Swiss National Science Foundation under contract PZ00P2\_154829. The work of A.F. and G.O. is supported in part by the National Science Foundation under Grant No. PHY-1417354.}

\bibliographystyle{JHEP}
\bibliography{mybib}

\providecommand{\href}[2]{#2}\begingroup\raggedright\begin{thebibliography}{10}

\bibitem{Cullen:2011ac}
G.~Cullen, N.~Greiner, G.~Heinrich, G.~Luisoni, P.~Mastrolia, G.~Ossola et~al.,
  \emph{{Automated One-Loop Calculations with GoSam}},
  \href{http://dx.doi.org/10.1140/epjc/s10052-012-1889-1}{\emph{Eur. Phys. J.}
  {\bf C72} (2012) 1889}, [\href{http://arxiv.org/abs/1111.2034}{{\tt
  1111.2034}}].

\bibitem{Cullen:2014yla}
G.~Cullen et~al., \emph{{G$\scriptsize{O}$S$\scriptsize{AM}$-2.0: a tool for
  automated one-loop calculations within the Standard Model and beyond}},
  \href{http://dx.doi.org/10.1140/epjc/s10052-014-3001-5}{\emph{Eur. Phys. J.}
  {\bf C74} (2014) 3001}, [\href{http://arxiv.org/abs/1404.7096}{{\tt
  1404.7096}}].

\bibitem{Broggio:2015lya}
A.~Broggio, A.~Ferroglia, B.~D. Pecjak, A.~Signer and L.~L. Yang,
  \emph{{Associated production of a top pair and a Higgs boson beyond NLO}},
  \href{http://dx.doi.org/10.1007/JHEP03(2016)124}{\emph{JHEP} {\bf 03} (2016)
  124}, [\href{http://arxiv.org/abs/1510.01914}{{\tt 1510.01914}}].

\bibitem{Broggio:2016zgg}
A.~Broggio, A.~Ferroglia, G.~Ossola and B.~D. Pecjak, \emph{{Associated
  production of a top pair and a W boson at next-to-next-to-leading logarithmic
  accuracy}}, \href{http://dx.doi.org/10.1007/JHEP09(2016)089}{\emph{JHEP} {\bf
  09} (2016) 089}, [\href{http://arxiv.org/abs/1607.05303}{{\tt 1607.05303}}].

\bibitem{Broggio:2016lfj}
A.~Broggio, A.~Ferroglia, B.~D. Pecjak and L.~L. Yang, \emph{{NNLL resummation
  for the associated production of a top pair and a Higgs boson at the LHC}},
  \href{http://dx.doi.org/10.1007/JHEP02(2017)126}{\emph{JHEP} {\bf 02} (2017)
  126}, [\href{http://arxiv.org/abs/1611.00049}{{\tt 1611.00049}}].

\bibitem{Broggio:2017kzi}
A.~Broggio, A.~Ferroglia, G.~Ossola, B.~D. Pecjak and R.~D. Sameshima,
  \emph{{Associated production of a top pair and a Z boson at the LHC to NNLL
  accuracy}}, \href{http://dx.doi.org/10.1007/JHEP04(2017)105}{\emph{JHEP} {\bf
  04} (2017) 105}, [\href{http://arxiv.org/abs/1702.00800}{{\tt 1702.00800}}].

\bibitem{Becher:2014oda}
T.~Becher, A.~Broggio and A.~Ferroglia, \emph{{Introduction to Soft-Collinear
  Effective Theory}}, vol.~896.
\newblock Springer, 2015,
  \href{http://dx.doi.org/10.1007/978-3-319-14848-9}{10.1007/978-3-319-14848-9}.

\bibitem{Ferroglia:2009ep}
A.~Ferroglia, M.~Neubert, B.~D. Pecjak and L.~L. Yang, \emph{{Two-loop
  divergences of scattering amplitudes with massive partons}},
  \href{http://dx.doi.org/10.1103/PhysRevLett.103.201601}{\emph{Phys. Rev.
  Lett.} {\bf 103} (2009) 201601}, [\href{http://arxiv.org/abs/0907.4791}{{\tt
  0907.4791}}].

\bibitem{Ferroglia:2009ii}
A.~Ferroglia, M.~Neubert, B.~D. Pecjak and L.~L. Yang, \emph{{Two-loop
  divergences of massive scattering amplitudes in non-abelian gauge theories}},
  \href{http://dx.doi.org/10.1088/1126-6708/2009/11/062}{\emph{JHEP} {\bf 11}
  (2009) 062}, [\href{http://arxiv.org/abs/0908.3676}{{\tt 0908.3676}}].

\bibitem{Nogueira:1991ex}
P.~Nogueira, \emph{{Automatic Feynman graph generation}},
  \href{http://dx.doi.org/10.1006/jcph.1993.1074}{\emph{J.Comput.Phys.} {\bf
  105} (1993) 279--289}.

\bibitem{Vermaseren:2000nd}
J.~A.~M. Vermaseren, \emph{{New features of FORM}},
  \href{http://arxiv.org/abs/math-ph/0010025}{{\tt math-ph/0010025}}.

\bibitem{Reiter:2009ts}
T.~Reiter, \emph{{Optimising Code Generation with haggies}},
  \href{http://dx.doi.org/10.1016/j.cpc.2010.01.012}{\emph{Comput.Phys.Commun.}
  {\bf 181} (2010) 1301--1331}, [\href{http://arxiv.org/abs/0907.3714}{{\tt
  0907.3714}}].

\bibitem{Cullen:2010jv}
G.~Cullen, M.~Koch-Janusz and T.~Reiter, \emph{{Spinney: A Form Library for
  Helicity Spinors}},
  \href{http://dx.doi.org/10.1016/j.cpc.2011.06.007}{\emph{Comput.Phys.Commun.}
  {\bf 182} (2011) 2368--2387}, [\href{http://arxiv.org/abs/1008.0803}{{\tt
  1008.0803}}].

\bibitem{Mastrolia:2012bu}
P.~Mastrolia, E.~Mirabella and T.~Peraro, \emph{{Integrand reduction of
  one-loop scattering amplitudes through Laurent series expansion}},
  \href{http://dx.doi.org/10.1007/JHEP11(2012)128,
  10.1007/JHEP06(2012)095}{\emph{JHEP} {\bf 06} (2012) 095},
  [\href{http://arxiv.org/abs/1203.0291}{{\tt 1203.0291}}].

\bibitem{vanDeurzen:2013saa}
H.~van Deurzen, G.~Luisoni, P.~Mastrolia, E.~Mirabella, G.~Ossola and
  T.~Peraro, \emph{{Multi-leg One-loop Massive Amplitudes from Integrand
  Reduction via Laurent Expansion}},
  \href{http://dx.doi.org/10.1007/JHEP03(2014)115}{\emph{JHEP} {\bf 03} (2014)
  115}, [\href{http://arxiv.org/abs/1312.6678}{{\tt 1312.6678}}].

\bibitem{Peraro:2014cba}
T.~Peraro, \emph{{Ninja: Automated Integrand Reduction via Laurent Expansion
  for One-Loop Amplitudes}},
  \href{http://dx.doi.org/10.1016/j.cpc.2014.06.017}{\emph{Comput. Phys.
  Commun.} {\bf 185} (2014) 2771--2797},
  [\href{http://arxiv.org/abs/1403.1229}{{\tt 1403.1229}}].

\bibitem{Ossola:2006us}
G.~Ossola, C.~G. Papadopoulos and R.~Pittau, \emph{{Reducing full one-loop
  amplitudes to scalar integrals at the integrand level}},
  \href{http://dx.doi.org/10.1016/j.nuclphysb.2006.11.012}{\emph{Nucl. Phys.}
  {\bf B763} (2007) 147--169}, [\href{http://arxiv.org/abs/hep-ph/0609007}{{\tt
  hep-ph/0609007}}].

\bibitem{Mastrolia:2008jb}
P.~Mastrolia, G.~Ossola, C.~Papadopoulos and R.~Pittau, \emph{{Optimizing the
  Reduction of One-Loop Amplitudes}},
  \href{http://dx.doi.org/10.1088/1126-6708/2008/06/030}{\emph{JHEP} {\bf 0806}
  (2008) 030}, [\href{http://arxiv.org/abs/0803.3964}{{\tt 0803.3964}}].

\bibitem{Ossola:2008xq}
G.~Ossola, C.~G. Papadopoulos and R.~Pittau, \emph{{On the Rational Terms of
  the one-loop amplitudes}},
  \href{http://dx.doi.org/10.1088/1126-6708/2008/05/004}{\emph{JHEP} {\bf 0805}
  (2008) 004}, [\href{http://arxiv.org/abs/0802.1876}{{\tt 0802.1876}}].

\bibitem{Mastrolia:2010nb}
P.~Mastrolia, G.~Ossola, T.~Reiter and F.~Tramontano, \emph{{Scattering
  AMplitudes from Unitarity-based Reduction Algorithm at the Integrand-level}},
  \href{http://dx.doi.org/10.1007/JHEP08(2010)080}{\emph{JHEP} {\bf 08} (2010)
  080}, [\href{http://arxiv.org/abs/1006.0710}{{\tt 1006.0710}}].

\bibitem{Binoth:2008uq}
T.~Binoth, J.~P. Guillet, G.~Heinrich, E.~Pilon and T.~Reiter, \emph{{Golem95:
  A Numerical program to calculate one-loop tensor integrals with up to six
  external legs}},
  \href{http://dx.doi.org/10.1016/j.cpc.2009.06.024}{\emph{Comput. Phys.
  Commun.} {\bf 180} (2009) 2317--2330},
  [\href{http://arxiv.org/abs/0810.0992}{{\tt 0810.0992}}].

\bibitem{Heinrich:2010ax}
G.~Heinrich, G.~Ossola, T.~Reiter and F.~Tramontano, \emph{{Tensorial
  Reconstruction at the Integrand Level}},
  \href{http://dx.doi.org/10.1007/JHEP10(2010)105}{\emph{JHEP} {\bf 1010}
  (2010) 105}, [\href{http://arxiv.org/abs/1008.2441}{{\tt 1008.2441}}].

\bibitem{Guillet:2013msa}
J.~P. Guillet, G.~Heinrich and J.~F. von Soden-Fraunhofen, \emph{{Tools for NLO
  automation: extension of the golem95C integral library}},
  \href{http://dx.doi.org/10.1016/j.cpc.2014.03.009}{\emph{Comput. Phys.
  Commun.} {\bf 185} (2014) 1828--1834},
  [\href{http://arxiv.org/abs/1312.3887}{{\tt 1312.3887}}].

\bibitem{CMS:2017uib}
{\scshape CMS} collaboration, C.~Collaboration, \emph{{Measurement of top
  pair-production in association with a W or Z boson in pp collisions at 13
  TeV}}, {\emph{CMS-PAS-TOP-17-005} (2017) }.

\bibitem{Aad:2015eua}
{\scshape ATLAS} collaboration, G.~Aad et~al., \emph{{Measurement of the $
  t\overline{t}W $ and $ t\overline{t}Z $ production cross sections in pp
  collisions at $ \sqrt{s}=8 $ TeV with the ATLAS detector}},
  \href{http://dx.doi.org/10.1007/JHEP11(2015)172}{\emph{JHEP} {\bf 11} (2015)
  172}, [\href{http://arxiv.org/abs/1509.05276}{{\tt 1509.05276}}].

\bibitem{CMS:2016dui}
{\scshape CMS} collaboration, C.~Collaboration, \emph{{Measurement of the top
  pair-production in association with a W or Z boson in $pp$ collisions at 13
  TeV}}, {\emph{CMS-PAS-TOP-16-017} (2016) }.

\bibitem{Alwall:2014hca}
J.~Alwall, R.~Frederix, S.~Frixione, V.~Hirschi, F.~Maltoni, O.~Mattelaer
  et~al., \emph{{The automated computation of tree-level and next-to-leading
  order differential cross sections, and their matching to parton shower
  simulations}}, \href{http://dx.doi.org/10.1007/JHEP07(2014)079}{\emph{JHEP}
  {\bf 07} (2014) 079}, [\href{http://arxiv.org/abs/1405.0301}{{\tt
  1405.0301}}].

\bibitem{Cascioli:2011va}
F.~Cascioli, P.~Maierhofer and S.~Pozzorini, \emph{{Scattering Amplitudes with
  Open Loops}},
  \href{http://dx.doi.org/10.1103/PhysRevLett.108.111601}{\emph{Phys. Rev.
  Lett.} {\bf 108} (2012) 111601}, [\href{http://arxiv.org/abs/1111.5206}{{\tt
  1111.5206}}].

\bibitem{Denner:2016kdg}
A.~Denner, S.~Dittmaier and L.~Hofer, \emph{{Collier: a fortran-based Complex
  One-Loop LIbrary in Extended Regularizations}},
  \href{http://dx.doi.org/10.1016/j.cpc.2016.10.013}{\emph{Comput. Phys.
  Commun.} {\bf 212} (2017) 220--238},
  [\href{http://arxiv.org/abs/1604.06792}{{\tt 1604.06792}}].

\bibitem{Zhang:2016kfo}
Y.~Zhang, \emph{{Lecture Notes on Multi-loop Integral Reduction and Applied
  Algebraic Geometry}},  2016.
\newblock \href{http://arxiv.org/abs/1612.02249}{{\tt 1612.02249}}.

\bibitem{Ita:2015tya}
H.~Ita, \emph{{Two-loop Integrand Decomposition into Master Integrals and
  Surface Terms}},
  \href{http://dx.doi.org/10.1103/PhysRevD.94.116015}{\emph{Phys. Rev.} {\bf
  D94} (2016) 116015}, [\href{http://arxiv.org/abs/1510.05626}{{\tt
  1510.05626}}].

\bibitem{Abreu:2017xsl}
S.~Abreu, F.~Febres~Cordero, H.~Ita, M.~Jaquier, B.~Page and M.~Zeng,
  \emph{{Two-Loop Four-Gluon Amplitudes with the Numerical Unitarity Method}},
  \href{http://arxiv.org/abs/1703.05273}{{\tt 1703.05273}}.

\bibitem{Mastrolia:2011pr}
P.~Mastrolia and G.~Ossola, \emph{{On the Integrand-Reduction Method for
  Two-Loop Scattering Amplitudes}},
  \href{http://dx.doi.org/10.1007/JHEP11(2011)014}{\emph{JHEP} {\bf 1111}
  (2011) 014}, [\href{http://arxiv.org/abs/1107.6041}{{\tt 1107.6041}}].

\bibitem{Badger:2012dp}
S.~Badger, H.~Frellesvig and Y.~Zhang, \emph{{Hepta-Cuts of Two-Loop Scattering
  Amplitudes}}, \href{http://dx.doi.org/10.1007/JHEP04(2012)055}{\emph{JHEP}
  {\bf 1204} (2012) 055}, [\href{http://arxiv.org/abs/1202.2019}{{\tt
  1202.2019}}].

\bibitem{Zhang:2012ce}
Y.~Zhang, \emph{{Integrand-Level Reduction of Loop Amplitudes by Computational
  Algebraic Geometry Methods}},
  \href{http://dx.doi.org/10.1007/JHEP09(2012)042}{\emph{JHEP} {\bf 1209}
  (2012) 042}, [\href{http://arxiv.org/abs/1205.5707}{{\tt 1205.5707}}].

\bibitem{Mastrolia:2012an}
P.~Mastrolia, E.~Mirabella, G.~Ossola and T.~Peraro, \emph{{Scattering
  Amplitudes from Multivariate Polynomial Division}},
  \href{http://dx.doi.org/10.1016/j.physletb.2012.09.053}{\emph{Phys.Lett.}
  {\bf B718} (2012) 173--177}, [\href{http://arxiv.org/abs/1205.7087}{{\tt
  1205.7087}}].

\bibitem{Mastrolia:2013kca}
P.~Mastrolia, E.~Mirabella, G.~Ossola and T.~Peraro, \emph{{Multiloop Integrand
  Reduction for Dimensionally Regulated Amplitudes}},
  \href{http://dx.doi.org/10.1016/j.physletb.2013.10.066}{\emph{Phys.Lett.}
  {\bf B727} (2013) 532--535}, [\href{http://arxiv.org/abs/1307.5832}{{\tt
  1307.5832}}].

\bibitem{Britto:2004nc}
R.~Britto, F.~Cachazo and B.~Feng, \emph{{Generalized unitarity and one-loop
  amplitudes in N = 4 super-Yang-Mills}},
  \href{http://dx.doi.org/10.1016/j.nuclphysb.2005.07.014}{\emph{Nucl. Phys.}
  {\bf B725} (2005) 275--305}, [\href{http://arxiv.org/abs/hep-th/0412103}{{\tt
  hep-th/0412103}}].

\bibitem{Mastrolia:2016dhn}
P.~Mastrolia, T.~Peraro and A.~Primo, \emph{{Adaptive Integrand Decomposition
  in parallel and orthogonal space}},
  \href{http://dx.doi.org/10.1007/JHEP08(2016)164}{\emph{JHEP} {\bf 08} (2016)
  164}, [\href{http://arxiv.org/abs/1605.03157}{{\tt 1605.03157}}].

\bibitem{vonManteuffel:2014ixa}
A.~von Manteuffel and R.~M. Schabinger, \emph{{A novel approach to integration
  by parts reduction}},
  \href{http://dx.doi.org/10.1016/j.physletb.2015.03.029}{\emph{Phys. Lett.}
  {\bf B744} (2015) 101--104}, [\href{http://arxiv.org/abs/1406.4513}{{\tt
  1406.4513}}].

\bibitem{Peraro:2016wsq}
T.~Peraro, \emph{{Scattering amplitudes over finite fields and multivariate
  functional reconstruction}},
  \href{http://dx.doi.org/10.1007/JHEP12(2016)030}{\emph{JHEP} {\bf 12} (2016)
  030}, [\href{http://arxiv.org/abs/1608.01902}{{\tt 1608.01902}}].

\bibitem{Maierhoefer:2017hyi}
P.~Maierhoefer, J.~Usovitsch and P.~Uwer, \emph{{Kira - A Feynman Integral
  Reduction Program}},  \href{http://arxiv.org/abs/1705.05610}{{\tt
  1705.05610}}.

\bibitem{Borowka:2016ehy}
S.~Borowka, N.~Greiner, G.~Heinrich, S.~Jones, M.~Kerner, J.~Schlenk et~al.,
  \emph{{Higgs Boson Pair Production in Gluon Fusion at Next-to-Leading Order
  with Full Top-Quark Mass Dependence}},
  \href{http://dx.doi.org/10.1103/PhysRevLett.117.012001}{\emph{Phys. Rev.
  Lett.} {\bf 117} (2016) 012001}, [\href{http://arxiv.org/abs/1604.06447}{{\tt
  1604.06447}}].

\bibitem{Borowka:2016ypz}
S.~Borowka, N.~Greiner, G.~Heinrich, S.~P. Jones, M.~Kerner, J.~Schlenk et~al.,
  \emph{{Full top quark mass dependence in Higgs boson pair production at
  NLO}}, \href{http://dx.doi.org/10.1007/JHEP10(2016)107}{\emph{JHEP} {\bf 10}
  (2016) 107}, [\href{http://arxiv.org/abs/1608.04798}{{\tt 1608.04798}}].

\bibitem{Heinrich:2017kxx}
G.~Heinrich, S.~P. Jones, M.~Kerner, G.~Luisoni and E.~Vryonidou, \emph{{NLO
  predictions for Higgs boson pair production with full top quark mass
  dependence matched to parton showers}},
  \href{http://dx.doi.org/10.1007/JHEP08(2017)088}{\emph{JHEP} {\bf 08} (2017)
  088}, [\href{http://arxiv.org/abs/1703.09252}{{\tt 1703.09252}}].

\end{thebibliography}\endgroup

%\begin{thebibliography}{99}
%\bibitem{...}
%....
%
%\end{thebibliography}

\end{document}